\documentclass[10pt,twocolumn,twoside]{IEEEtran}
\usepackage{cite}
\usepackage{amssymb}
\usepackage{float}
\usepackage{eqparbox}
\usepackage{multirow}

\usepackage[cmex10]{amsmath}
\usepackage{amsthm}

\usepackage{algorithm}
\usepackage{algorithmic}

\usepackage{array}

\usepackage{perpage} 
\MakePerPage{footnote} 

\usepackage{mdwmath}
\usepackage{mdwtab}
\usepackage{amsmath} \usepackage{amsfonts}
\makeatletter \def\amsbb{\use@mathgroup \M@U \symAMSb} \makeatother
\usepackage{bbold}
\usepackage{subfigure}
\usepackage[belowskip=-12pt,aboveskip=0pt]{caption}
\usepackage{cite}
\usepackage{graphicx}
\usepackage{booktabs} 
\usepackage{lipsum}
\usepackage{amssymb}
\usepackage{amsthm}
\usepackage{eqparbox}
\usepackage{multirow}
\usepackage[cmex10]{amsmath}
\usepackage{stfloats}  

\begin{document}
\title{Centralized Adaptation for Parameter Estimation over Wireless Sensor Networks}
\author{Reza~Abdolee,~\IEEEmembership{Student Member,~IEEE,}
       and~  Benoit~Champagne,~\IEEEmembership{Senior Member,~IEEE}

\thanks{R. Abdolee and B. Champagne are with the Department
of Electrical and Computer Engineering, McGill University, Montreal, QC H3A 0E9, Canada (e-mail: reza.abdolee@mail.mcgill.ca, benoit.champagne@mcgill.ca).}

}

\maketitle
\def\H{\mbox{\boldmath $H$}}
\def\A{\mbox{\boldmath $A$}}
\def\D{\mbox{\boldmath $D$}}
\def\E{\amsbb{E}}
\def\Y{\boldmath{Y}}
\def\X{\boldmath{X}}
\def\bphi{\mbox{\boldmath $\phi$}}
\def\bpsi{\mbox{\boldmath $\psi$}}
\def\btheta{\mbox{\boldmath $\theta$}}
\def\bbeta{\mbox{\boldmath $\beta$}}
\def\blambda{\mbox{\boldmath $\lambda$}}
\def\balpha{\mbox{\boldmath $\alpha$}}
\def\bgamma{\mbox{\boldmath $\gamma$}}
\def\bPsi{\mbox{\boldmath $\Psi$}}
\def\bomega{\mbox{\boldmath $\omega$}}
\def\bsigma{\mbox{\boldmath $\sigma$}}

\def\diag{\mbox{\rm{diag}}}
\def\col{\mbox{\rm{col}}}
\def\msd{{\rm{msd}}}
\def\emse{{\rm{emse}}}
\def\for{{\;\rm{for}\;}}
\def\bvec{{\rm{bvec}}}
\def\vec{{\rm{vec}}}
\def\var{{\rm{var}}}
\def\Ind{{\rm{Ind}}}
\def\rank{{\rm{rank}}}
\def\net{{\rm{net}}}
\def\Tr{{\rm{Tr}}}
\def\n{\boldsymbol{n}}
\def\g{\boldsymbol{g}}
\def\Ical{\boldsymbol{\cal I}}
\def\a{\boldsymbol{a}}
\def\e{\boldsymbol{e}}
\def\f{\boldsymbol{f}}
\def\h{\boldsymbol{h}}
\def\d{\boldsymbol{d}}
\def\p{\boldsymbol{p}}
\def\t{\boldsymbol{t}}
\def\u{\boldsymbol{u}}
\def\v{\boldsymbol{v}}
\def\w{\boldsymbol{w}}
\def\x{\boldsymbol{x}}
\def\y{\boldsymbol{y}}
\def\z{\boldsymbol{z}}
\def\be{\begin{equation}}
\def\ee{\end{equation}}
\def\ba{\begin{align}}
\def\ea{\end{align}}
\begin{abstract}
We study the performance of centralized least mean-squares (CLMS) algorithms in wireless sensor networks where nodes transmit their data over fading channels to a central processing unit (e.g., fusion center or cluster head), for parameter estimation. Wireless channel impairments, including fading and path loss, distort the transmitted data, cause link failure and  degrade the performance of the adaptive solutions.
To address this problem, we propose a novel CLMS algorithm that uses a refined version of the transmitted data and benefits from a link failure alarm strategy to discard severely distorted data. Furthermore, to remove the bias due to communication noise from the estimate, we introduce a bias-elimination scheme that also leads to a lower steady-state mean-square error.
Our theoretical findings are supported by numerical simulation results.
\end{abstract}
\section{Introduction}
\label{sec:intro}
\IEEEPARstart{C}{entralized} least mean-squares (CLMS) algorithms were
introduced in previous studies for applications in networks with multiple processing nodes \cite{lopes2008diffusion,cattivelli2010diffusion}. In the development of these algorithms, it is commonly assumed that the communication links between the nodes and the fusion center are error free, i.e., there is no channel distortion such as fading, path loss, and noise during the data transmission. Consequently, these algorithms may experience substantial performance degradation when used over wireless sensor networks (WSN), as evidenced in recent studies on distributed estimation in WSN operating over fading channels\cite{abdolee2011diffusion,abdolee2013diffusion}.  

Reference  \cite{abdolee2011diffusion}  and \cite{abdolee2013diffusion} propose novel distributed adaptive algorithms for networks with imperfect communication links. However, these algorithms do not include provisions to handle possible link failures as well as the estimation bias resulting from the link noise.
Work in  \cite{bertranddiffusion2011} studies the bias compensation problem for scenarios where only the regression data are noisy. Previous work \cite{abdolee2013estimation}  investigates the problem of space-varying parameter estimation over networks using distributed processing approaches. In these works, however, the communication links between the nodes are assumed to be ideal, i.e., no fading and communication noise are taken into consideration.

We, in this paper, bring all these factors into account and study the performance of CLMS algorithms for parameter estimation in WSN in the presence of radio channel impairments.
Specifically, we propose a novel CLMS algorithm that uses a refined version of the transmitted information and benefits from a link failure alarm strategy to discard severely distorted data.
We demonstrate the presence of a bias in the adaptive parameter estimates, consequence of the link noise, and introduce a bias-elimination scheme that also significantly decreases the steady-state mean-square error (MSE) of the network. Beside the application of our proposed bias-removal technique in centralized networks, it can be also used in networks with a distributed hierarchical structure, where the cluster heads act as fusion centers and distributed processing is performed at a cluster level. 

The adaptive feature of the proposed algorithm is highly desirable in non-stationary signal environments where the underlying network  parameters change over time.
In summary, the contribution of this letter are as follow: a) development of a new CLMS algorithm for parameter estimation in WSN operating over fading channels; b) analysis of the developed algorithm and identification of the main technical issue under such condition, namely the estimation bias; c) derivation of the relationship between the bias estimates with that of the channel perturbations; d) development of a new bias-compensation technique to remove the bias  and e) performance analysis of the developed bias-compensated algorithm, including the derivation  of the stability conditions and the steady-state mean-square deviation (MSD) expressions.

\section{Problem Formulation}
\label{sec.:ProblemFormulation}
Consider a network of  $N$ randomly distributed sensor nodes that are deployed over a geographical area to estimate an unknown parameter vector $w^o \in {\amsbb C}^{M\times 1}$.
At time instant $i \in \amsbb{N}$, node $k \in \{1,2,\cdots,N\}$ collects a measurement datum, $\d_k(i)\in {\amsbb C}$, that is related to $w^o$ through a linear model\footnote{In this letter, the same mathematical notation as in  \cite{abdolee2013estimation} is used, where boldface fonts are reserved for random variables.}, i.e.:
\vspace{-0.15cm}
\be
\d_k(i)=\u_{k,i} w^o+\v_k(i)
\label{eq.:network-data-model}
\ee
where the regression vectors $\u_{k,i}\in {\amsbb C}^{1 \times M}$ are zero-mean  i.i.d. in time and independent over space with positive definite covariance matrix $R_{u,k}=\E [\u_{k,i}^* \u_{k,i}]$,  the noise terms $\v_k(i) \in {\amsbb C} $  are zero-mean  i.i.d. in time and independent over space with variance $\sigma^2_{v,k}$.
The regression data $\u_{k,i}$ and the measurement noise $\v_{\ell}(j)$ are independent  \cite{lopes2008diffusion,cattivelli2010diffusion,zhao2012performance}.

Let us assume that each node $k$ assembles its available data into a row vector as $\z_{k,i}=[\u_{k,i}, \d_k(i)]$ and sends it over a wireless channel to the network fusion center.
During its transmission, the data experience distortion of the form:
\vspace{-0.15cm}
\be
{\hat \z}_{k,i} =\a_{k}(i) \z_{k,i}+\v^{(z)}_{k,i}
\label{eq.:received_distorted_data}
\ee
where ${\hat \z }_{k,i}=[{\hat \u}_{k,i},\,  {\hat \d}_{k}(i)]$ denotes the received data at the fusion center and $\v^{(z)}_{k,i} \in {\amsbb C}^{1 \times (M+1)}$ is the link noise with $\v^{(z)}_{k,i}=[\v^{(u)}_{k,i},\v^{(d)}_k(i)]$.
We consider   $\a_{k}(i)=\h_{k}(i)\sqrt{\frac{P}{r_{k}^{\alpha}}}$ for analog  data transmission, where $\h_{k}(i) \in {\amsbb C}$ represents the fading channel coefficient, $P$ is the transmit signal power, $r_{k}$ is the distance between node $k$ and the fusion center and $\alpha$ is the path loss exponent.  For digital-type data communication, the coefficient $\a_{k}(i)$ will be an indicator function that models the link-failure (packet loss) in baseband after decoding. The fading coefficients $\h_{k}(i)$ are zero-mean circular complex Gaussian, i.i.d. in time\footnote{This assumption is true when the time interval between successive iterations of the adaptive process is larger than the coherence time of the channels.} and independent over space with variance $\sigma^2_{h,k}$.
The link noise vectors $\v^{(z)}_{k,i}$ are zero-mean, i.i.d. in time and independent over space with covariance matrix $\sigma^{(z)2}_{v,k}I$, while the components $\v^{(u)}_{k,i}$ and $\v^{(d)}_k(i)$ are independent.   The random variables $\h_{k}(i)$,  $\u_{k,i}$,  $\v_k(i)$ and $\v^{(z)}_{k,i}$ are all mutually independent. We use ${r^o_k}$  to denote  the free-space transmission range of node $k$, as obtained using Friis formula for a the given power budget, antenna gains and carrier frequency \cite{nikitin2006theory}.
Furthermore, we let $\varsigma^o_{k}$ represent the threshold signal-to-noise ratio (SNR) of the received signal from node $k$, defined as the SNR of the received signal over a non-fading link with communication range $r^o_k$, i.e., $
\varsigma^o_{k}={{P}/{\sigma^{(z)2}_{v,k} (r_k^o)^{\,\alpha}}}
$.
We  use $\boldsymbol{\varsigma}_{k}(i)$ to denote the link instantaneous SNR, which is given by $
\boldsymbol{\varsigma}_{k}(i)={{|
\h_{k}(i)|^2P}/{\sigma^{(z)2}_{v,k}r_{k}^{\alpha}}}$.  The transmission from node $k$ to the fusion center will be successful if $\boldsymbol{\varsigma}_k(i)$ exceeds the threshold level $\varsigma_{k}^o$, i.e., if $
|\h_{k}(i)|^2\geq ({r_{k}}/{r^o_k})^\alpha
$; otherwise, the link fails. In the digital case, this amounts to a packet loss, while in the analog case this would correspond to a temporary loss of synchronization. 
Since the channel coefficients, $\h_k(i)$, are zero mean circular complex Gaussian random variables, their squared magnitudes, $|\h_k(i)|^2$, have exponential distribution  \cite{Garcia1994probability}.
Let $\lambda=1/\sigma^2_{h,k}$, then the probability of successful transmission will be:
\begin{align}
 p_k&=\textrm{Pr}\Big(|\h_{k}(i)|^2\geq \big(\frac{ r_{k}}{r_k^o}\big)^{\alpha}\Big)
=e^{-\lambda(r_{k f}/{r^o_k})^{\alpha}}
\label{eq.:probability-of-sucess}
\end{align}
\section{Centralized LMS (CLMS) Algorithm}
\label{sec.:CentralizedLMSAlgorithms}
Let us first consider the analog-type data communication between nodes and the fusion center.  In this case, the fusion center pre-process $\{{\hat \u}_{k,i}, {\hat \d}_k(i)\}_{k=1}^N$ to partially recover $\{\u_{k,i},\d_k(i)\}_{k=1}^N$ before using them in the LMS iteration. This pre-processing can be of linear form, e.g.:
\vspace{-0.2cm}
\begin{align}
&{\bar{\u}}_{k,i}\triangleq \g_{k}(i) {\hat \u}_{k,i}
\label{eq.:equalized-received_regressor}\\
&{\bar{\d}}_{k}(i)\triangleq \g_{k}(i){\hat \d}_k(i)
\label{eq.:equalized-received_d}
\end{align}
where $\g_{k}(i)$ are scalar equalization coefficients. Assuming negligible channel estimation error, these coefficients can be obtained, in practice by, e.g.,  the least squares (LS) method:
\vspace{-0.15cm}
\be
\g_{k}(i)=\frac{\h_{k}^*(i)}{|\h_{k}(i)|^2} \sqrt{\frac{r_k^{\alpha}}{P}}
\label{eq.equalization-entries}
\ee
Substituting (\ref{eq.equalization-entries}) into (\ref{eq.:equalized-received_regressor}) and (\ref{eq.:equalized-received_d}) leads to:
\vspace{-0.2cm}
\begin{align}
&{\bar{\u}}_{k,i}={\u}_{k,i}+\g_{k}(i)\v^{(u)}_{k,i}
\label{eq.:zero-forcing-equalized-received_regressor}\\
&{\bar{\d}}_{k}(i)={\d}_{k}(i)+\g_{k}(i)\v^{(d)}_{k}(i)
\label{eq.:zero-forcing-equalized-received_d}
\end{align}
The network can now seek the unknown parameter vector $w^o$ using the pre-processed data by minimizing:
\vspace{-0.2cm}
\be
\bar{J}(w)=\sum_{k=1}^N \E\Big[\Ical_k (i) \big|{\bar{\d}}_{k}(i)-{\bar{\u}}_{k,i}w\big|^2\Big]
\label{eq.ctrlObjFunctionFading}
\ee
where $\Ical_k (i)$ is a random variable with Bernouilli distribution defined as:
\vspace{-0.2cm}
\begin{align}
\Ical_k (i)=
\left\{\begin{array}{l}
1,\quad {\rm if} \; \boldsymbol{\varsigma}_{k}(i) \geq \varsigma^o_k \\
0, \quad \; {\rm otherwise}
\end{array}
\label{eq.:Ical-CtrlLms}
\right.
\end{align}
That is, $\Ical_{k}(i)=1$ when the transmission from node $k$ to the fusion center is successful and $\Ical_{k}(i)=0$ otherwise. We note that
the probability of success, i.e., $p_k=\E[\Ical_k (i)]$, is given by (\ref{eq.:probability-of-sucess}).
Since the cost function (\ref{eq.ctrlObjFunctionFading}) is strictly convex, its optimal point will be its only stationary point. This leads to \cite{sayed2008}:
\vspace{-0.15cm}
\be
w^o_{\textrm{ctrl}}=\Big(\sum_{k=1}^N \big(p_k R_{u,k}+R_{v,k}\big )\Big)^{-1}\Big(\sum_{k=1}^N p_k \, r_{du,k}\Big )
\label{eq.ctrlMMSEsolutionLinkFading}
\ee
where $ r_{du,k}=\E[\d_k(i) \u^*_{k,i}]$, and
\begin{align}
R_{v,k}&= \E\big[\Ical_k (i) |\g_{k}(i)|^2\big] R_{v,k}^{(u)}
\label{eq.:Rvuh}
\end{align}
with $R_{v,k}^{(u)}=\E[\v^{(u)*}_{k,i} \v^{(u)}_{k,i}]$.
The expectation term on the right hand side of (\ref{eq.:Rvuh}) can be obtained as:
\vspace{-0.15cm}
\begin{align}
\E\big[\Ical_k (i) |\g_{k}(i)|^2\big]&= \int_{0^+}^{y_{k}^o}  \frac{r_k^\alpha}{P\, y_k} \lambda e^{-\lambda \big( \frac{r_k^\alpha}{P\, y_k}-x_{k}^o\big)} dy_k
\label{eq.:Ey}
\end{align}
where $y_{k}^o\triangleq\frac{({r^o_k})^{\alpha}}{P }$ and $x_{k}^o \triangleq (\frac{r_k}{r^o_k})^{\alpha}$.

Since the covariance and cross-covariance of the data may not be available in practice, the estimate $w^o_{\textrm{ctrl}}$ given by (\ref{eq.ctrlMMSEsolutionLinkFading}) can alternatively be sought using the following  CLMS algorithm:
\vspace{-0.2cm}
\begin{align}
\w_i=\w_{i-1}+\mu \sum_{k=1}^N \Ical_k (i) {\bar{\u}}_{k,i}^{\ast} \big({\bar{\d}}_{k}(i)-{\bar{\u}}_{k,i} \w_{i-1}\big)
\label{alg.:centralized_lms_fading}
\end{align}
where $\mu>0$ is the step-size. The CLMS algorithm for digital data transmission model, i.e., when $\a_k(i)=\Ical_k (i)$ will be the same except that ${\bar{\u}}_{k,i}$ and ${\bar{\d}}_{k}(i)$ are now replaced by ${\hat{\u}}_{k,i}$ and ${\hat{\d}}_{k}(i)$, respectively. In this case,  $\Ical_k (i) $ will be an indicator function that models the link-failure or packet loss due to fading. We now proceed to investigate the convergence of the CLMS algorithm. To simplify the presentation, we first consider the case $\a_k(i)=1$ and assume that the links fail with probability $p_k$.
Later, we include fading and path loss effects into the analysis.
Under this condition, from  (\ref{eq.:received_distorted_data}), we have:
\vspace{-0.2cm}
\begin{align}
{\hat \d}_k(i)&=\d_{k}(i)+\v^{(d)}_k(i) \label{eq:dkf-noisy}\\
{\hat \u}_{k,i}&=\u_{k,i}+\v^{(u)}_{k,i} \label{eq:ukf-noisy}
\end{align}
Using (\ref{eq.:network-data-model}), (\ref{eq:dkf-noisy}) and (\ref{eq:ukf-noisy}),  we then obtain:
\vspace{-0.15cm}
\begin{align}
&{\hat \d}_k(i)={\hat \u}_{k,i} w^o+\hat{\v}_k(i)
\label{eq.:d_noisy_with_wo} \\
& {\hat \v}_k(i)=\v_k(i)+\v_k^{(d)}(i) -\v^{(u)}_{k,i} w^o
\end{align}
The variance of $ {\hat \v}_k(i)$ is then given by $
\sigma^2_{\hat{v},k}=\sigma^2_{v,k}+\sigma^{(d)2}_{v,k}+w^{o*} R_{v,k}^{(u)} w^o
$. We now define the weight error-vector $\tilde{\w}_i=w^o-\w_i$ and use (\ref{alg.:centralized_lms_fading}) and (\ref{eq.:d_noisy_with_wo}) to arrive at:
\vspace{-0.2cm}
\begin{align}
\tilde{\w}_i= \D_i \tilde{\w}_{i-1}-\mu  \t_i
\label{eq.:CtrlErrorVectorBiased}
\end{align}
where $ \D_i \triangleq I-\mu \sum_{k=1}^N \Ical_k (i) \hat{\u}^*_{k,i} {\hat \u}_{k,i}$ and $\t_i \triangleq \sum_{k=1}^N \Ical_k (i) \hat{\u}^*_{k,i} {\hat \v}_k(i)$.  Taking expectation from both sides of (\ref{eq.:CtrlErrorVectorBiased}), we obtain:
\vspace{-0.2cm}
\begin{align}
\E[\tilde{\w}_i]=D \E [\tilde{\w}_{i-1}]-\mu\, t_{f}
\label{eq.:ErrorVectorRecursionBiased}
\end{align}
where $D \triangleq \E[\D_i]=I-\mu \sum_{k=1}^N p_k\big(R_{u,k}+R_{v,k}^{(u)}\big)$ and $t_{f} \triangleq \E[ \t_i]=-\sum_{k=1}^N p_k R_{v,k}^{(u)} w^o$.
From recursion (\ref{eq.:ErrorVectorRecursionBiased}), it can be verified that the  CLMS algorithm will be mean-stable if
\vspace{-0.2cm}
\be
0<\mu<\frac{2}{ \lambda_{\text{max}}\Big(\sum_{k=1}^N p_k\big(R_{u,k}+R_{v,k}^{(u)})\Big)}
\label{eq.CtrlLmsRangeLinkFailure}
\ee
If the step-size satisfy (\ref{eq.CtrlLmsRangeLinkFailure}), from (\ref{eq.:ErrorVectorRecursionBiased}), we can also show that the mean estimate of the algorithm deviates from the optimal estimate by:
\vspace{-0.5cm}
\begin{align}
b& \triangleq w^o-\lim_{i \rightarrow \infty}\E[\w_i]=\Big(\sum_{k=1}^N p_k\big(R_{u,k}+R_{v,k}^{(u)} \big)\Big)^{-1} t_{f}
\label{eq.CtrlLmsRangeLinkFaiureBias}
\end{align}
We observe that the bias, $b$, does not depend on the step-size $\mu$. Therefore, reducing the step-size will not improve the accuracy of the estimated parameters.

To incorporate the effects of fading and path-loss into the above results, we can use expressions (\ref{eq.:zero-forcing-equalized-received_regressor}) and (\ref{eq.:zero-forcing-equalized-received_d}) and repeat the analysis. Doing so, we obtain similar results as before, except that in this case the power of the communication noises $\v^{(u)}_{k,i}$ and $\v^{(d)}_{k}(i)$  increase by the factor of $\E[\Ical_k (i) |\g_{k}(i)|^2]$. Consequently,  (\ref{eq.CtrlLmsRangeLinkFailure}) and (\ref{eq.CtrlLmsRangeLinkFaiureBias}) can still be used if,  in these expressions, we replace $p_k R_{v,k}^{(u)}$ with $R_{v,k}$.

\section{Bias-Compensated CLMS}
\label{sec.:biasCompenstaedCLmsFadingChannel}
As seen from (\ref{eq.CtrlLmsRangeLinkFaiureBias}), the mean of the weight error vector of the CLMS algorithm in (\ref{alg.:centralized_lms_fading}),  i.e., $\E[{\tilde \w}_i]$, converges to
a non-zero vector. In what follows, we propose a bias-compensation scheme and develop a new form of centralized LMS algorithm whose mean estimation error converges to zero.

As shown in (\ref{eq.CtrlLmsRangeLinkFaiureBias}), the estimation bias $b$ is due to the term $t_f$, which in turn is caused by the regression noise component $\v^{(u)}_{k,i}$.
Assuming zero regression noise in (\ref{eq.ctrlObjFunctionFading}), the unbiased optimal estimate of the network will be:
\vspace{-0.2cm}
\be
w^o=\Big(\sum_{k=1}^N p_k R_{u,k} \Big)^{-1}\Big(\sum_{k=1}^N p_k \, r_{du,k}\Big )
\label{eq.UnbiasedMMSEsolutionLinkFading}
\ee
We, in this paper,  propose a CLMS algorithm that runs over the noisy received data and achieves the optimal estimate (\ref{eq.UnbiasedMMSEsolutionLinkFading}).
The basic idea in our development is to construct an objective function whose gradient vector is identical to that of the cost (\ref{eq.ctrlObjFunctionFading}) with zero regression noise \cite{abdoleeDiffNoisyRegreesor2012}. The following objective function satisfies this criterion:
\vspace{-0.2cm}
\be
J(w)=\sum_{k=1}^N \E\big[\Ical_k (i)\big(|{\bar{\d}}_{k}(i)-{\bar{\u}}_{k,i}w|^2-\|R_{v,k}^{ 1/2}w\|^2\big)\big]
\label{eq.ModifiedctrlObjFunctionFading}
\ee
It can be verified that the Hessian of  $J(w)$ is positive definite and hence this cost function is strongly convex.
From (\ref{eq.ModifiedctrlObjFunctionFading}), we then arrive at the following bias-compensated CLMS (BC-CLMS) algorithm where ${\bar \mu}$ is the new step-size:
\vspace{-0.2cm}
\begin{align}
\w_i&=\w_{i-1}+{\bar \mu} \sum_{k=1}^N \Ical_k (i)\Big \{ {\bar{\u}}_{k,i}^{\ast} \big({\bar{\d}}_{k}(i)-{\bar{\u}}_{k,i} \w_{i-1})\nonumber \\
&\hspace{4cm}+\g_k(i)R_{v,k}^{(u)} \w_{i-1}\Big\}
\label{alg.:bias_compensated_centralized_lms_fading}
\end{align}
This algorithm requires the covariance matrix of the regression noise, which can be estimated off-line or during the operation of the algorithm in real-time as in e.g., \cite{zheng2003least,bertranddiffusion2011}.
The bias-compensated CLMS algorithm for digital data transmission will be the same if we replace ${\bar{\u}}_{k,i}$ and ${\bar{\d}}_{k}(i)$ with ${\hat{\u}}_{k,i}$ and ${\hat{\d}}_{k}(i)$, respectively and set $\g_k(i)$ to one.

We now proceed to analyze the proposed algorithm. Similar to previous section, we first consider the case with $\a_k(i)=1$.
Under this condition, it can be verified that the weight error vector of the algorithm evolves with time according to:
\vspace{-0.2cm}
\begin{align}
\tilde{\w}_i= {\bar \D}_i \tilde{\w}_{i-1}-\mu  {\bar \t}_i
\label{eq.:CtrlErrorVector}
\end{align}
where, in this case, $
{\bar \D}_i  \triangleq I-\mu \sum_{k=1}^N \Ical_k (i) \big({\hat \u}^*_{k,i}  {\hat \u}_{k,i} -R_{v,k}^{(u)}\big) $ and
$ {\bar \t}_i \triangleq \sum_{k=1}^N \Ical_k (i) \big({\hat \u}^*_{k,i}  \hat{\v}_k(i)+R_{v,k}^{(u)}w^o) $.
By taking the expectation of (\ref{eq.:CtrlErrorVector}), we arrive at:
\begin{align}
\E[\tilde{\w}_i]={\bar D}\, \E [\tilde{\w}_{i-1}]
\label{eq.:BiasCompensatedCtrlErrorVector}
\end{align}
where $
{\bar D} \triangleq \E[{\bar \D}_i]=I-\mu \sum_{k=1}^N p_k R_{u,k}
$. It can be verified that $\lim_{i \rightarrow \infty} \E[\tilde{\w}_i] \rightarrow 0$ if
\vspace{-0.2cm}
\be
0<{\bar \mu}<\frac{2}{ \lambda_{\text{max}}\Big(\sum_{k=1}^N p_k R_{u,k}\Big)}
\label{eq.BiasCompensatedCtrlLmsRangeLinkFailure}
\ee
The mean-square deviation (MSD) recursion of this algorithm can be expressed as:
\vspace{-0.2cm}
\be
\E\|\tilde \w_i\|^2_{\Sigma}=\E\|\tilde \w_{i-1}\|^2_{\Sigma'}+\big(\vec\big({\bar \mu}^2 {\bar X})\big)^* \vec(\Sigma)
\ee
where $\E \| \tilde \w_i \|^2_{\Sigma}=\E \|w^o-\w_i \|^2_{\Sigma}$,  $\Sigma\geq 0$, $ \Sigma' \triangleq \E[{\bar \D}_i \Sigma {\bar \D}_i]$ and
\vspace{-0.25cm}
\begin{align}
&{\bar X}=\sum_{k=1}^N p_k \Big (\sigma^2_{v,kf} R_{u,k}+(\beta-1)\, (R_{v,k}^{(u)} w^o w^{o*} R_{v,k}^{(u)}) \nonumber \\
&\hspace{3cm}+R_{v,k}^{(u)}\Tr(w^o w^{o*} R_{v,k}^{(u)})\Big)
\end{align}
Using this recursion, the steady-state MSD of the algorithm converges to:
\begin{align}
{\bar \eta}_{\text{ctrl}}= \big[\vec ({\bar \mu}^2 {\bar X})\big]^* (I-{\bar F})^{-1}\, \vec(I_M)
\label{eq.:msd-bias-compensatedCtrlLms}
\end{align}
where ${\bar F}\approx {\bar D}^T \otimes {\bar D}^* $. As mentioned before,  the steady-state MSD of the network under fading condition can be computed by replacing $p_k R_{v,k}^{(u)}$ with $R_{v,k}$ given by (\ref{eq.:Rvuh}).
\section{Simulation Results}
\label{sec.:CtrlLmsFadingResults}
\begin{figure}
\centering
\includegraphics[width=7.5cm ,height=5.5cm]{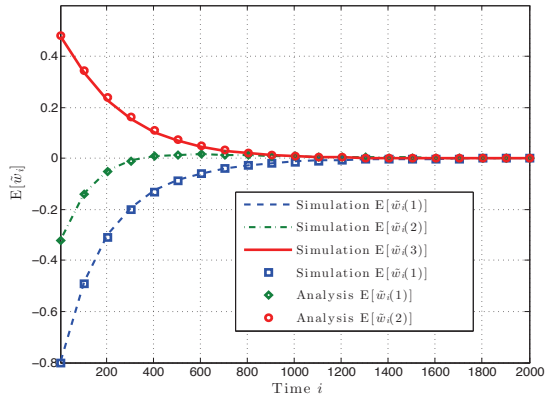}
\caption{\small{Mean error vector versus time for (\ref{alg.:bias_compensated_centralized_lms_fading}).}}
\label{fig.:WeightErrorVectorUnbiasedTVC}
\end{figure}

In this section, we present computer experiments to illustrate the performance of the CLMS algorithm (\ref{alg.:centralized_lms_fading}) and the BC-CLMS algorithm (\ref{alg.:bias_compensated_centralized_lms_fading}) over a wireless sensor network. 
We consider a WSN consisting of $N=5$ nodes, uniformly distributed over a squared area of $[x,y]=[1,1]$ km with the fusion center located at $x=y=0.5$km. The unknown parameter vector to be estimated is $w^o=[-0.8006,-0.3203+j0.1601,0.4804]^T$ with $M=3$.
All nodes have equal transmit signal power $P=10$mW,
the nominal transmission range is $r^o_k=0.3$km,
and the path loss exponent is $\alpha=2.5$.
We initialize the algorithms with $\w_{-1}=[0, 0, 0]^T$ and set the step-size to $\mu=0.003$.
To generate communication noise vectors $\v^{(z)}_{k,i}$,
we adopt complex normal distributions with zero mean and variances $\sigma^{(z)2}_{v,k} \in \{ 0.0617,    0.0560,    0.0923,    0.0831,    0.0476\}$.
The measurement noise $\v_k(i)$ is generated using a normal distribution with variances $\sigma^2_{v,k} \in \{0.069,    0.090,    0.087,    0.092,    0.061,\}$.
The fading coefficients of the radio links are obtained from a circular complex normal distribution with zero mean and unit variance.
The regression data $\u_{k,i}$ are generated from a circular complex Gaussian distribution where the $(n,m)$-th entry of correlation matrix $R_{u,k}$ is given by $\eta_k^{m-n}\, \for m\geq n$, with $\eta_k=k/(2N)$.
%
Figure \ref{fig.:WeightErrorVectorUnbiasedTVC} shows the mean convergence performance of the BC-CLMS algorithm  (\ref{alg.:bias_compensated_centralized_lms_fading}). In this figure, we illustrate  the convergence of the real part of the mean error vector $\E[\tilde{\w}_i]$. As the results show the mean error vector converges to zero after 1200 iterations. In contrast, the mean error vector of the CLMS algorithm (\ref{alg.:centralized_lms_fading}) converges to the biased value of $b= -0.1489 - j0.0086, 
  -0.0278 + j0.0317, 0.1064 - j0.0025]^T $,  which is in agreement with our analytical finding. 
We examine the mean-square performance of CLMS  and BC-CLMS algorithms for two cases. In the first case, the fading coefficients change at each time $i$ and links fail according to the channels instantaneous SNR. In our results, we refer to this case as the fading case. For the second case, the communication links are distorted by additive noise but links fail with probability of $1-p_k$ with $p_k=[0.342, 0.336, 0.415, 0.363, 0.473]$. We refer to the second case as the link-failure case.
\begin{figure}
\centering
\includegraphics[width=7.5cm ,height=5.5cm]{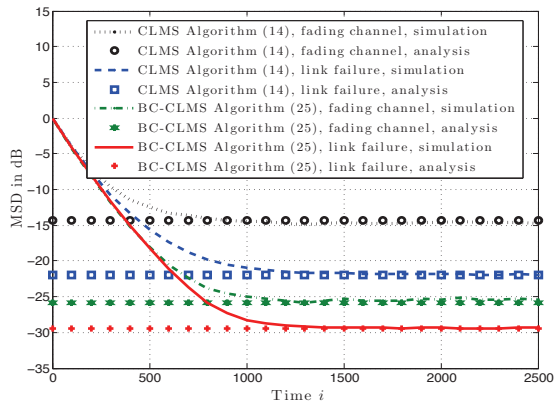}
\caption{\small{MSD versus time for CLMS (14) and BC-CLMS (25).}}
\label{fig.:TransientMsdBiasedAndUnbiased}
\end{figure}
\begin{figure}
\centering
\includegraphics[width=7.5cm ,height=5.5cm]{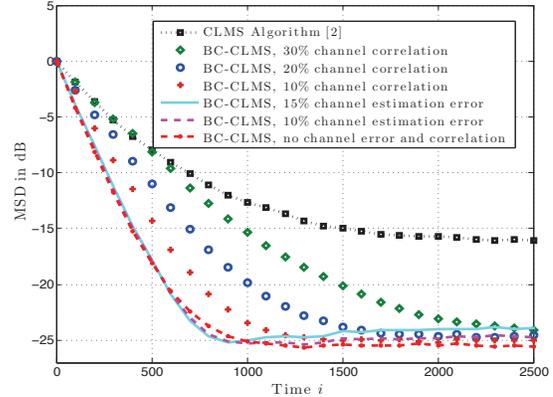}
\caption{\small{The MSD performance of BC-CLMS algorithm under channel estimation error and temporal correlation.}}
\label{fig.:TransientMsdChanError}
\end{figure}
Figure \ref{fig.:TransientMsdBiasedAndUnbiased} shows the MSD of the algorithms for these two cases. As expected, the BC-CLMS algorithm achieves a significantly lower steady-state MSD than the CLMS in both cases.   In Fig. \ref{fig.:TransientMsdChanError}, we compare the performance of the proposed BC-CLMS algorithm with that of the conventional CLMS algorithm proposed in \cite{cattivelli2010diffusion}. As this figure shows the proposed algorithm outperforms CLMS with about 8dB. This figure also shows the performance of the BC-CLMS algorithm under imperfect channel information at the fusion center and under temporal channel correlation.  We observe that as channel correlation increases the converges speed of the algorithm decreases while its steady-state performance remains invariant. In contrast, as the channel estimation error increases the steady-state MSD of the algorithm increases. As this result indicates, the proposed algorithm is robust to small channel estimation error and temporal correlation.
\section{Conclusion}
\label{sec.:conclusion}
We studied the performance of CLMS algorithms for parameter estimation over wireless sensor networks where links between nodes and the fusion center are impaired by fading and noise.
The analysis and the numerical simulations show that the proposed BC-CLMS algorithm is stable and converges to an unbiased estimate for small adaptation step-sizes.
\typeout{}

\end{document}